\documentclass{ws-ijgmmp}

\usepackage{comment}
\usepackage{hhline}
\usepackage{empheq}

\DeclareMathOperator{\Christ}{\Gamma}
\newcommand{\STChrist}{\overset{\rm st}{\Gamma}{}}
\newcommand{\affChrist}{\overset{\rm aff}{\Gamma}{}}

\newcommand{\alphanabla}{\nabla^{(\alpha)}}

\newcommand{\LCChrist}{\overset{\rm{\scriptscriptstyle LC}}{\Gamma}{}}

\begin{document}

\markboth{T. Wada}
{Reframing of IG via STG}

%
\catchline{}{}{}{}{}
%

\title{REFRAMING OF INFORMATION GEOMETRY VIA SYMMETRIC TELEPARALLEL GRAVITY
}

\author{TATSUAKI WADA 
}

\address{Region of Electrical and Electronic Systems Engineering, Ibaraki University, \\
4-12-1 Nakanarusawa, Hitachi, 316-8511, Japan\,
\\
\email{tatsuaki.wada.to@vc.ibaraki.ac.jp}
 }

\maketitle

\begin{history}
\received{(Day Month Year)}
\revised{(Day Month Year)}
\end{history}

\begin{abstract}
Information geometry has traditionally been formulated within the framework of Riemannian geometry and dual affine connections. This work aims to reframe the foundational structure of information geometry by introducing the geometric machinery of symmetric teleparallel gravity. By requiring both curvature and torsion to vanish globally on the statistical manifold, it is demonstrated that the fundamental properties of the information space can be entirely encoded in the non-metricity tensor.

This approach clarifies the distinction between the general $\xi $-parameterized case and the $\theta$- or $\eta $-parameterized cases, which correspond to general symmetric teleparallel gravity and its coincident gauge, respectively. Specifically, the $\theta$- or $\eta $-coordinates emerge as special coordinates in the coincident gauge, where the connection coefficients vanish.
\end{abstract}


\section{Introduction}
\label{intro}

Recently, insights from the holographic principle and relativistic quantum information suggest that gravitational phenomena and emergent geometry can be effectively reformulated through information-theoretic measures. Within this context, information geometry (IG)  \cite{AN00, Amari}  provides a powerful mathematical framework. By employing the Fisher information metric and the dual connections, IG allows us to map the statistical properties of informational or thermodynamic states directly onto geometric manifolds.
However, conventional attempts to synthesize IG with gravitational physics typically rely on Einstein's General Relativity (GR), where gravity is mediated by spacetime curvature. This approach often encounters conceptual friction, as the dually flat structures inherent in IG do not naturally align with the metric-compatible, torsion-free Riemann-Cartan manifold of GR. To resolve this tension, it is highly advantageous to shift the geometric paradigm from curvature to non-metricity. Specifically, Symmetric Teleparallel Gravity (STG) \cite{NY98, A06,H24,JHK18,BGGM24,T25} offers an alternative framework where both curvature and torsion vanish, and the gravitational interaction is fully encoded in the non-metricity tensor.
STG is an alternative way to explain gravity. Unlike the standard GR, it has no curvature and no torsion. Instead, it uses a flat framework called non-metricity.
Nowadays, it is known that gravity can be described in three different ways. Each way uses a different geometrical object.
Together, these are so called the \textit{geometric trinity} \cite{JHK19,CCF25}.
\begin{itemize}
 \item GR: spacetime has curvature (torsion and non-metricity are zero).
  \item Teleparallel Gravity: spacetime has torsion (curvature and non-metricity are zero).
 \item STG: spacetime is completely flat (curvature and torsion are zero), but it has non-metricity.
\end{itemize}
Non-metricity tensor
\begin{align}
 Q_{kij} := \nabla_k g_{ij} = \partial_k g_{ij} - \Christ^{\ell}{}_{ki} g_{\ell j} - \Christ^{\ell}{}_{kj} g_{i \ell},
  \label{Qcf}
\end{align}
characterizes the deviation from the metric condition $Q_{kij} = \nabla_k g_{ij} = 0$.
In general, if the metric $g$ and the connection $\nabla$ are chosen arbitrarily, then $Q$ is not zero and the non-metricity $Q \neq 0$ holds.
In other words, since the metric $g$, the connection $\nabla$, and the non-metricity $Q$ must satisfy \eqref{Qcf}, two of these three can be chosen freely.
If we impose the metric condition $Q=0$, then once we fix one of the remaining two (either the metric $g$ or the connection $\nabla$) the other is determined.

In STG, spacetime is flat and parallel lines remain parallel globally. However, the metric $g$ varies along the manifold. Non-metricity $Q := \nabla g$ characterizes this spatial variation of the measurement tool. Crucially, the dynamics of $Q$ generate the geometric effects that are macroscopically perceived as gravity. 
Furthermore, one can always choose a special coordinate system, say $\{ \phi^a \}_{a=1}^n$, in which all components of the affine connection $\nabla$ vanish globally. This choice is referred to as the coincident gauge \cite{BGGM24,JK22,BL22}, which significantly simplifies the complex gravitational field equations.

On the other hand,
information geometry (IG) \cite{AN00, Amari} is a powerful framework that applies the techniques of differential geometry to the field of information sciences. It treats probability distributions as points on a curved surface known as a \textit{statistical manifold} $\mathcal{S}$.
Dually flat spaces are essential in IG. A dually flat space is a Riemannian manifold equipped with a pair of mutually dual affine connections $\nabla$ and $\nabla ^{\star} $ that are both flat. This mathematical framework is highly important because it bridges differential geometry and mathematical statistics, turning complex probabilistic optimization into intuitive geometric problems.
While much research has focused on the mathematical foundations of IG, studies on its physical foundation are limited.
Caticha \cite{C15,C16} has developed IG as a foundation for physics, particularly in the fields of statistical inference and thermodynamics. 
Since both curvature and torsion vanish in dually flat spaces, non-metricity appears to be the remaining geometric quantity responsible for their nontrivial structure. Nevertheless, to the best of my knowledge, the term "non-metricity" is rarely used in the IG literature.
Together with my collaborators, the gradient-flow \cite{N94,FA95} in IG had been studied from the different perspectives and are related to some different fields of physics. Ref. \cite{CW24} studied the gradient-flow from the perspective of
analytical mechanics and  applied to  black hole thermodynamics. 
Through the study \cite{W23, WN25} of IG from the perspective of Weyl's gauge symmetry, the importance of non-metricity in dually-flat spaces has been revealed.
The importance of non-metricity in IG was recently studied in \cite{WS26} and showed 
the  non-metricity tensor associated with the $\alpha$-connection $ \nabla^{ ( \alpha )}$ as
\begin{align}
  \alphanabla_i \; g_{jk} = \alpha \, C_{ijk},
 \label{Qalpha}
\end{align}
where $C_{ijk}$ denote Amari-Centsov tensor.
Furthermore, the $\alpha$-connection is derived \cite{WS26} based on the non-metricity tensor characterized by Amari-Centsov tensor.

As observed, STG and IG have a lot in common.
Table \ref{tab} highlights the geometric similarities shared between STG and IG.
In particular, the correspondence between the coincident gauge in STG and the dually flat space in IG plays a crucial role.
\begin{table}[h]
\label{tab}
\caption{Comparison of STG and the dually flat spaces in IG.}
\label{tab1}%
\centering
\begin{tabular}{| l | l | l  | l |} 
\hline
geometric quantity & STG   & $\nabla$-flat  & $\nabla^{\star}$-flat  \\
\hhline{| = | = | = | = |}
affine connection    & $ \STChrist^k{}_{ij}(\xi) $  & $\overset{\theta}{\Gamma}{}^k{}_{ij}(\xi)$
& $\overset{\eta}{\Gamma}{}^k{}_{ij}(\xi)$ \\
&  & $e$-connection & $m$-connection  \\
\hline 
curvature    & 0    & 0  & 0 \\
\hline
torsion   & 0  & 0  & 0  \\
\hline
non-mentricity   & $\overset{\rm st}{\nabla}_k g_{ij}(\xi) \ne 0$   & $\nabla_k g_{ij}(\xi) \ne 0$ & $\nabla^{\star}{}_k g^{ij}(\xi) \ne 0$ \\
\hline
special coordinates   & $\{ \phi^a \}_{a=1}^n$  & $\{ \theta^a \}_{a=1}^n$ & $\{ \eta_a \}_{a=1}^n$\\
in coincident gauge &  $\STChrist^c{}_{ab}(\phi) = 0$   & $\overset{\theta}{\Gamma}{}^c{}_{ab}(\theta) = 0$
& $\overset{\eta}{\Gamma}{}_c{}^{ab}(\eta) = 0$   \\
\hline
\end{tabular}
\end{table}
In this contribution, I shall reframe the physical foundational structure of IG by introducing the geometric machinery of STG. By enforcing both curvature and torsion to vanish globally on the statistical manifold $\mathcal{S}$, we demonstrate that the fundamental properties of information space can be entirely encoded into the non-metricity tensor in the coincident gauge. 
In IG the so called \textit{exponential family} \eqref{exp-pdf} is a premier example which provides the theoretical foundation of  a dually-flat space.
IG has traditionally been formulated within the framework of Riemannian geometry endowed with the Fisher metric $g^{\rm F}$ and the dual affine connections $\nabla$ and $\nabla^{\star}$ on the statistical manifold $\mathcal{S}$.
In contrast, the reframing allows us to distinguish the general $\xi$-parameterized case from the $\theta$- or $\eta$-parameterized cases, mirroring the relationship between general and the coincident gauge cases in STG. The $\theta$- or $\eta$-coordinates in the dually flat spaces emerge as the special coordinates in the coincident gauge, where the connection coefficients vanish globally.

The next section provides some preliminaries: the geometric features in STG in 2.1 and
the review on the basics of IG in 2.2.
Section \ref{reframing} reframes IG via the geometric machinery of STG.
The St\"ckelberg trick is used to restore the covariance.
The covariant affine connections and the special coordinates systems in the coincident gauge play
key roles.  These geometrical machinery in STG allows us to distinguish the general $\xi$-parameterized case from the $\theta$- or $\eta$-parameterized case. 
Some examples for Gaussian model are given in 3.1.
Final section \ref{conclusion} is devoted to the conclusion.  \ref{app1} explains the St\"uckelberg  trick,
which restore covariance or invariance.  \ref{app2} provides a simple proof of the useful identity.

Through out the paper, we use the index notations familiar in general relativity.
Einstein summation convention is used.

\section{Preliminaries}
\subsection{Geometric features in STG }
Here we briefly review the geometric features in STG, which are necessary in this work.

The general coefficients $\affChrist^k{}_{ij}$ of a torsion-free affine connection can be decomposed into two parts \cite{BL22},
\begin{align}
  \affChrist^k{}_{ij} = \LCChrist^k{}_{ij} + L^k{}_{ij}.
\end{align} 
Here $\LCChrist^k{}_{ij}$ are the coefficients of Levi-Civita connection of the metric $g_{ij}$ and
\begin{align}
L^k{}_{ij} = \frac{1}{2} \, g^{k \ell} ( Q_{\ell i j} - Q_{i \ell j} - Q_{j \ell i} ) = L^k{}_{ji},
\end{align}
is a disformation tensor, which is constructed from the non-metricity tensors $Q_{kij}:= \nabla_k g_{ij}$.
A distinctive feature in STG is that one can find a specific coordinate system where all coefficients of the torsion-free affine connection vanishes globally on the manifold. Choosing this specific coordinate system is what we refer to as fixing the coincident gauge.
The non-metricity tensor $\nabla_k g_{ij}$ reduces to the simple form of $\partial_k g_{ij}$ in this coincident gauge.

Let us consider a manifold $\mathcal{M}$ covered by the coordinates $\{ \xi^i \}_{i=1}^n$ endowed with a metric $g_{ij}(\xi)$ and an affine connection $\nabla$
defining the notion of parallel transport on $\mathcal{M}$.
In STG, the covariant connection coefficients are in the form \cite{BGGM24}
\begin{align}
  \STChrist^k{}_{ij}(\xi) = \frac{\partial \xi^k}{\partial \phi^a} \frac{\partial^2 \phi^a}{\partial \xi^i \partial \xi^j},
  \label{STconnect}
\end{align}
which can be obtained by a coordinate transformation from a frame specified by $\{ \phi^a \}_{a=1}^n$ with vanishing
coefficients of the connection. The connection given by \eqref{STconnect} is flat and is symmetric in the lower indices, i.e., torsion-free.
It is worth noting that
if $ \phi^a $ transforms as a vector, Eq. \eqref{STconnect} is not covariant (not symmetric under the diffeomorphisms). 
The appropriate transformation for $\phi^a$ is treating them as a set of scalar fields $\phi^a(\xi)$,
which can be regarded as the St\"uckelberg fields (see  \ref{app1}).
The set of the scalar fields $\{ \phi^a(\xi) \}_{a=1}^n$ represent the coordinate system of the coincident gauge in which $\STChrist^c{}_{ab}(\phi) = 0$.

\subsection{Basics of Information Geometry}
Here we review the conventional framework   \cite{AN00,Amari}  of IG.

A probability distribution function (pdf) belongs to the exponential family, if it is written in the form
\begin{align}
  p_{\theta}(x) = \exp \big[ \theta^a F_a(x) - \Psi(\theta)  \big],
  \label{exp-pdf}
\end{align}
which is parametrized by $\{ \theta^a \}_{a=1}^n$.
Here $x$ denotes a value of a stochastic variable, each $F_a(x)$ is a function of $x$, and
the $\theta$-potential $\Psi(\theta)$ is determined from the normalization of $p_{\theta}(x)$ as $\Psi(\theta) = \ln \left[ \int dx \exp (\theta^a F_a(x) ) \right]$, i.e.,
the cumulant generating function. 

A dually flat structure is a foundational concept in IG, where a smooth manifold is simultaneously equipped with a Riemannian metric and a pair of distinct, flat affine connections that are mutually dual with respect to the metric. Formally, a dually flat structure is defined as a quadruple $(\mathcal{S}, g, \nabla, \nabla^{\star})$.
Here $\mathcal{S}$ is a statistical manifold, whose underlying space representing a $n$-parameter space $\{ \theta^a\}_{a=1}^n$ of an exponential pdf \eqref{exp-pdf},
$g$ is the Fisher metric, the primal connection $\nabla$ is a torsion-free affine connection whose curvature tensor is zero (flat),
and its dual connection $\nabla ^{\star}$ is a second torsion-free, flat affine connection. 
The two connections $\nabla$ and $\nabla^{\star}$ are related by the celebrated formula
\begin{align}
  \partial_a g_{bc}= \Gamma^d{}_{ab} \; g_{d c} + \Gamma^{\star}{}^d{}_{ac} \; g_{b d}.
  \label{dualC}
 \end{align}
Because both connections are flat, the manifold $S$ admits two independent, global coordinate systems:
the coordinates $ \{ \theta^a \}_{a=1}^n$ are flat with respect to $\nabla$ and 
the dual coordinates $ \{ \eta_a \}_{a=1}^n$ are flat with respect to $\nabla ^{\star}$.
The basis $\partial_a := \partial / \partial \theta^a$ and the dual basis $\partial^a := \partial / \partial \eta_a$ are
$\nabla$-parallel and $\nabla^{\star}$-parallel, i.e.,
\begin{align}
  \nabla_{\partial_a} \partial_b = 0, \quad \nabla_{\partial^a} \partial^b=0,
\end{align}
respectively. Both bases are biorthogonal, i.e., $\langle \partial_a, \partial^b \rangle = \delta^b_a$, where $\delta^b_a$ denotes Kronecker's delta.

The geometry is completely determined by the pair of convex potential functions $\psi(\theta)$ and $\psi^{\star}(\eta)$ which are linked through the Legendre transformation
\begin{align}
 \psi^{\star} (\eta ) =\sup _{\theta } \{\theta^a \, \eta_a -\psi (\theta ) \}.
 \end{align}
Here   $\Psi^\star(\eta)$ is the (negative sign of) entropic function ${\rm E}_{p_{\theta}} [ \ln  p_{\theta}(\xi) ]$,
where ${\rm E}_{p_{\theta}} [ \cdot ]$ denotes the expectation with respect to $p_{\theta}(x)$.
The coordinates are gradients of these potentials
 \begin{align}
  \eta _a = \frac{\partial \Psi^{\star}(\eta)}{ \partial \theta ^a}, \quad \textrm{and} \quad
  \theta ^a = \frac{\partial \Psi(\theta)}{\partial \eta _a}.
   \end{align}
 The metric tensor can be calculated directly by taking the Hessian of either potential
 \begin{align} 
g_{ab} ( \theta) = 
\frac{\partial^2 \Psi( \theta)}{\partial \theta^a \partial \theta^b}, \quad
g^{ab} ( \eta) = 
\frac{\partial^2 \Psi^\star(\eta)}{\partial \eta_a \partial \eta_b}.
\label{g} 
\end{align}
They satisfy the relation $g^{ab} (\eta) \, g_{bc} (\theta) = \delta^a_c$.
This dually flat structure, formulated within the framework of pure differential geometry, is equivalent to what is known as a Hessian structure in Hessian geometry \cite{Shima}.

The difference between any two pdfs, say $p_{\theta}(x)$ and $q_{\theta}(x)$, is measured by Kullback-Leibler (KL) divergence
\begin{align}
 D_{\rm KL}(p, q) &:= \int_{-\infty}^{\infty} dx \; p_{\theta}(x) \ln \frac{p_{\theta}(x)}{q_{\theta}(x)} 
 = {\rm E}_{p_{\theta}} \left[ \ln \frac{p_{\theta}(x)}{q_{\theta}(x)} \right], 
 \label{KL}
\end{align}
which is a non-negative function providing a measure how much they differ.
When both pdfs belong to the exponential family \eqref{exp-pdf}, the KL divergence \eqref{KL} is equivalent to
\begin{align}
  D(\eta^p \vert \eta^q) & = 
 \Psi^{\star}(\eta^p) - \Psi^{\star}(\eta^q) - \theta_q^i ( \eta_i^p - \eta_i^q),
\end{align}
which is the Bregman divergence with respect to a convex function of $\Psi^{\star}(\eta)$.
It is worth noting that in dually flat spaces, the metric is not used to measure distance. Instead, the KL (or Bregman) divergence is employed.

Having explained the geometric features in ST gravity (or geometry) and the basics of IG, we shall
reframing IG in the next section.

\section{Reframing IG via STG}
\label{reframing}

Here IG is reframed via the geometric machinery of STG and the St\"uckelberg  trick.

Let $p_{\xi}(\xi)$ denotes a pdf parametrized by a set of parameters $\{ \xi^i \}_{i=1}^n$.
This $p_{\xi}(\xi)$ characterizes a statistical model in IG.
Recall that the Fisher metric with respect to $p_{\xi}(x)$ is given by
\begin{align}
   {\rm E}_{p_{\xi}} \left[ \frac{\partial \ln p_{\xi}(x)}{\partial \xi^i}  \frac{\partial \ln p_{\xi}(x)}{\partial \xi^j} \right],
   \label{Fm}
\end{align}
which is not the Hessian of a convex function in general.
However, in the special case that  the pdf $p_{\xi}(\xi)$  belongs to the exponential family \eqref{exp-pdf}, i.e., $p_{\xi}(\xi)$
is cast into the form of $p_{\theta}(x)$, 
it is possible to induce the statistical manifold $\mathcal{S}$ with a Hessian metric $g(\theta)$, i.e., a Riemannian metric given by the potential $\Psi(\theta)$ of a convex function. In this case, the manifold naturally inherits two flat affine connections $\nabla$ and $\nabla^{\star}$. 
Not all but many pdfs belong to the exponential family \eqref{exp-pdf}. A well known example is Gaussian or normal pdf $N(\mu, \sigma^2)$, where $\mu$ and $\sigma^2$ are expectation and variance, respectively.
As a result, each $\theta^a$ is expressed as a function $ \theta^a(\xi)$ of the parameters $\{ \xi^i \}_{i=1}^n$.

Now let $\mathcal{M}$ be a manifold representing the parameter space with respect to the pdf $p_{\xi}(\xi)$
and  let $\{ \xi^i \}_{ i=1}^n$
be local coordinates on $\mathcal{M}$.
Following the St\"uckelberg  trick, we introduce a set of $n$ real scalar fields $ \{ \theta^a(\xi) \}_{a=1}^n$,
which are determined by the exponential family \eqref{exp-pdf} and play  a role as a coordinate system in the internal space $\mathcal{S}$ (or moving frame).
Here and hereafter in order to distinguish a vector or tensor in $\mathcal{M}$ from that in $\mathcal{S}$,
we use indices from the middle Latin alphabet (e.g., $i, j, k, ...$) for the former and indices from the beginning of Latin alphabet (e.g., $a, b, c, ...$) for the latter.
Unlike the conventional formulation of IG, we regard the dually flat connection as the two different symmetric teleparallel (ST) connections: one $\nabla$ is flat in the coincident gauge of $ \{ \theta^a(\xi) \}_{a=1}^n$; and the other $\nabla^{\star}$ is flat in the coincident gauge
of $ \{ \eta_a(\xi) \}_{a=1}^n$.

We denotes $\overset{ \theta}{ \Christ}{}^k{}_{ij}( \xi)$ as the coefficients of the ST connection $\nabla$ 
on $\mathcal{M}$ with respect to the St\"uckelberg fields $ \{ \theta^a(\xi) \}_{a=1}^n$, i.e., they satisfy
\begin{align}
  \nabla_{\partial_i} \; \partial_j =\overset{ \theta}{ \Christ}{}^k{}_{ij}( \xi) \; \partial_k.
\end{align}
Here each basis $\partial_a$ is related to the bases $\{ \partial_i := \partial / \partial \xi^i \}_{i=1}^n$ by
\begin{align}
  \partial_a = \frac{\partial \xi^i}{\partial \theta^a} \, \partial_i.
\end{align}
Since the $\theta$-coordinates are affine, all coefficients $\overset{\theta}{\Gamma}{}^c{}_{ab}(\theta)$ of the connection $\nabla$ vanish, i.e., $\nabla_{\partial_a} \partial_b = \overset{\theta}{\Gamma}{}^c{}_{ab}(\theta) \, \partial_c = 0$. Then it follows that
\begin{align}
 \overset{ \theta}{ \Christ}{}^k{}_{ij}( \xi) & \, \partial_k =  \nabla_{\partial_i} \underbrace{\partial_j}_{ (\partial \theta^b / \partial  \xi^j ) \partial_b} 
 =  \nabla_{\partial_i} \Big( \frac{ \partial\theta^b}{\partial \xi^j } \partial_b \Big) 
 =\Big(   \partial_i \frac{ \partial\theta^b}{\partial \xi^j }\Big)  \underbrace{\partial_b}_{(\partial \xi^k / \partial \theta^b ) \partial_k}  + \frac{ \partial\theta^b}{\partial \xi^j } \nabla_{\partial_i} \partial_b  \notag \\
 &=   \frac{\partial^2 \theta^b}{\partial \xi^i \partial \xi^j} \frac{\partial \xi^k}{ \partial \theta^b } \partial_k +  \frac{\partial \theta^b}{\partial \xi^j} \frac{\partial \theta^a}{\partial \xi^i} \underbrace{ \nabla_{\partial_a} \partial_b}_{0},
\end{align}
from which we obtain
\begin{align}
 \overset{ \theta}{ \Christ}{}^k{}_{ij}( \xi) =  \frac{\partial \xi^k}{\partial \theta^a} \;  \frac{\partial^2 \theta^a}{\partial \xi^i \partial \xi^j}.
 \label{thetaSTC}
\end{align}
The coefficients $ \overset{ \theta}{ \Christ}{}^k{}_{ij}( \xi)$ take the same form in the expression of the covariant coefficients of the ST connection \eqref{STconnect}.
Hence we see that the scalar fields $\{ \theta^a(\xi) \}_{a=1}^n$ represent the special coordinate system in the coincident gauge.
In a similar way, we can obtain the coefficients of the ST connection $\nabla^{\star}$ as
\begin{align}
 \overset{ \eta}{ \Christ}{}^k{}_{ij}( \xi) =  \frac{\partial \xi^k}{\partial \eta_a} \;  \frac{\partial^2 \eta_a}{\partial \xi^i \partial \xi^j},
 \label{etaSTC}
\end{align}
which are the other (dual) covariant coefficients of the ST connection.
The scalar fields  $\{ \eta_a(\xi) \}_{a=1}^n$ represent the other special coordinate system in the coincident gauge.

Notably, while conventional IG treats $\{ \theta^a \}_{a=1}^n$ and $\{ \eta_a \}_{a=1}^n$ as affine coordinates on the same statistical manifold, this reframing distinguishes the general coordinates $\{ \xi^i \}_{i=1}^n$ on $\mathcal{M}$ from these two distinct internal coordinate systems. Consequently, the dually flat spaces are realized as two different moving frames, $\{ \theta^a(\xi) \}_{a=1}^n$ and $\{ \eta_a(\xi) \}_{a=1}^n$, constructed over $\mathcal{M}$.

Next we reconsider the fact that the metric tensor is obtained by taking the Hessian of either potential as shown in \eqref{g}.
In IG \cite{Amari}, it is assumed that the dual affine connections are torsion-free.
Then, from the formula \eqref{formula} in \ref{app2}, we obtain $Q_{abc}(\theta) - Q_{cba}(\theta)=0$, i.e.,
\begin{align}
  Q_{abc}(\theta) = \nabla_a g_{bc}(\theta) =  \nabla_c g_{ba}(\theta) = Q_{cba}(\theta).
\end{align}
Combined with the symmetry of  the metric tensor, the non-metricity tensor $Q_{abc}(\theta)$ in IG is found to be totally symmetric under the exchange of indices.
In the $\theta$-coordinates of the coincident gauge, the totally symmetry of the non-metricity tensor $Q_{abc}(\theta)$ leads to
\begin{align}
   \partial_a g_{bc}(\theta) = \partial_b g_{ac}(\theta).
\end{align}
This is the necessary and sufficient condition that the metric $g_{ab}(\theta)$ is expressed as the Hessian
as shown in \eqref{g}.
In a similar way, the totally symmetry of $Q^{\star}{}^{abc}(\eta)$ leads to the latter relation in \eqref{g}.

Table \ref{tab2} summarizes the relationship between the geometries represented in the $\xi$-coordinates on $\mathcal{M}$ and the geometry represented in the $\theta$ frame of the flat space. 
Similar relationship holds for $\eta$-moving frame of the dual flat space.
These representations can be regarded as the information-geometric analogues of GR and STG, respectively. 
Through this analogy, the foundational structure in IG is reframed by utilizing the geometric machinery of STG and the St\"uckelberg trick, thereby restoring diffeomorphism invariance.

\begin{table}[h]
\caption{Comparison of the geometries of $\xi$-coordinates and $\theta$-frame. Curvature $R$, Torsion $T$, and non-metricity $Q$.
}\label{tab2}
\centering
\renewcommand{\arraystretch}{1.5}
 \begin{tabular}{|c|c|c|}
 \hline
  {rep.} & $R, T, Q$ & {conn. coeffs. (gravity theory) }\\
   \hhline{|=|=|=|}
  $\xi$-coord. & curved $R \ne 0, T = Q = 0$ & $\LCChrist{}^k{}_{ij}( \xi)$ (GR) \\
 \hline
   $\xi$-coord. & flat $R = T = 0, Q = \nabla g \ne 0$ & $ \overset{ \theta}{ \Christ}{}^k{}_{ij}( \xi) \ne 0$  (STG) \\
 \hline
  \multicolumn{3}{c}{ $\Downarrow \quad$ coincident gauge of $\{ \theta^a(\xi) \}_{a=1}^n$}
  \\   \hline
  $\theta$-frame & flat $R =  T  = 0, Q = \partial_a g_{bc}(\theta) \ne 0 $ & $ \overset{ \theta}{ \Christ}{}^c{}_{ab}( \theta) = 0$ (STG) \\
  \hline
 \end{tabular}
\end{table}

It is worthwhile to note that
while Hessian geometry \cite{Shima} and the conventional IG are restricted to the dually flat spaces governed by Hessian metrics \eqref{g}, this reframing enables the description of flat structures in the general $\xi$-coordinate  with respect to a pdf $p_{\xi}(x)$, which do not admit a Hessian metric in general.

\subsection{Gaussian model}
\label{Gaussian}
As an example, we here consider the Gaussian, or Normal $N(\mu, \sigma^2)$, pdf which is given by
\begin{align}
 p_{\xi}(x) = \frac{1}{\sqrt{2 \pi \, \sigma^2}} \,
     \exp \left[ -\frac{(x-\mu)^2}{2 \sigma^2} \right].
     \label{pxi}
\end{align}
Here $\mu$ denotes the mean, $\sigma$ is the standard deviation.
The $\xi$-coordinates are given by $(\xi^1, \xi^2) = (\mu, \sigma)$.
The non-zero components of the Fisher metric \eqref{Fm} are
\begin{align}
  g_{\mu \mu} = \frac{1}{\sigma^2}, \quad g_{\sigma \sigma} = \frac{2}{\sigma^2}.
  \label{gxi}
\end{align}
The corresponding non-zero components of Levi-Civita connection are
\begin{align}
  \LCChrist^{\mu}{}_{\mu \sigma} = \LCChrist^{\mu}{}_{\sigma \mu}  = -\frac{1}{\sigma},\quad
  \LCChrist^{\sigma}{}_{\mu \mu} = \frac{1}{2 \sigma}, \quad \LCChrist^{\sigma}{}_{\sigma \sigma} = -\frac{1}{\sigma}.
\end{align}
It is known \cite{AN00} that this is a hyperbolic space with the constant negative scalar curvature of $R=-1$.

Next we focus on the ST connections.
The non-zero components of the metric tensor are same as in \eqref{gxi} and the Fisher metric is not Hessian.
By expanding the $\xi$-parametrized pdf in \eqref{pxi}, it is cast into the form of the exponential family \eqref{exp-pdf} as follows.
\begin{align}
 p_{\xi}(x)  = \exp \Big[\underbrace{ \frac{\mu}{\sigma^2}}_{\theta^1} x \underbrace{- \frac{1}{2 \sigma^2}}_{\theta^2} x^2 - \Big(\underbrace{ \frac{\mu^2}{2 \sigma^2} + \frac{1}{2} \ln (2 \pi \sigma^2)}_{\Psi(\theta)} \Big) \Big]
 = p_{\theta}(x).
\end{align}
Then the St\"uckelberg fields $\{ \theta^a \}_{a=1}^2$ and  $\{ \eta_a \}_{a=1}^2$  are obtained as
\begin{align}
 \theta^1 = \frac{\mu}{\sigma^2}, \quad \theta^2 = -\frac{1}{2 \sigma^2}, \quad
 \eta_1 = \mu, \quad \eta_2 = \mu^2 + \sigma^2.
\end{align}
Straight forward calculations of \eqref{thetaSTC} and \eqref{etaSTC} lead to
\begin{align}
 \overset{ \theta}{ \Christ}{}^{\mu}{}_{\mu \sigma}( \xi)  =  \overset{ \theta}{ \Christ}{}^{\mu}{}_{\sigma \mu}( \xi)  = - \frac{2}{\sigma}, \quad \overset{ \theta}{ \Christ}{}^{\sigma}{}_{\sigma \sigma}( \xi) = -\frac{3}{\sigma}, \quad
  \overset{ \eta}{ \Christ}{}^{\sigma}{}_{\mu \mu}( \xi)  = \frac{1}{\sigma} \quad 
   \overset{ \eta}{ \Christ}{}^{\sigma}{}_{\sigma \sigma}( \xi) = \frac{1}{\sigma},
\end{align}
where the other components are zero.
The non-zero components of the non-metricity tensor \eqref{Qcf} are
\begin{align}
\overset{\theta}{Q}{}_{\mu \mu \sigma}(\xi) &= \overset{\theta}{Q}{}_{\mu \sigma \mu}(\xi) = \overset{\theta}{Q}{}_{\sigma \mu \mu}(\xi) = \frac{2}{\sigma^3}, \quad \overset{\theta}{Q}{}_{\sigma \sigma \sigma}(\xi) = \frac{8}{\sigma^3}, \notag \\
\overset{\eta}{Q}{}_{\mu \mu \sigma}(\xi) &= \overset{\eta}{Q}{}_{\mu \sigma \mu}(\xi) = \overset{\eta}{Q}{}_{\sigma \mu \mu}(\xi) = -\frac{2}{\sigma^3}, \quad \overset{\eta}{Q}{}_{\sigma \sigma \sigma}(\xi) = -\frac{8}{\sigma^3}.
\end{align}
The autoparallel equations \cite{WS26} for an arbitrary parameter, say $\lambda$, are
\begin{align}
  \frac{d  u^i(\lambda)}{d \lambda} + \STChrist^i{}_{jk}(\xi) u^j(\lambda) u^k(\lambda)
  = f(\lambda) u^i(\lambda), 
     \label{arbAPE}
  \end{align}
 with
 \begin{align}
  f(\lambda) = 
  \frac{1}{2 u^2(\lambda)} \left( \frac{d u^2(\lambda)}{d \lambda}-
  \overset{\textrm{st}}{Q}_{j k \ell}(\xi) u^j(\lambda) u^k(\lambda) u^{\ell}(\lambda) \right).
\end{align}
Here we introduced 
\begin{align}
 u^i(\lambda):= \frac{d \xi^i}{d \lambda}, \quad u^2(\lambda) := g_{jk}(\xi) \frac{d \xi^j}{d \lambda} \frac{d \xi^k}{d \lambda}.\end{align}
The connection coefficients $\STChrist^i{}_{jk}(\xi)$ are given by either $\overset{ \theta}{ \Christ}{}^{\mu}{}_{\sigma \mu}( \xi)$ or  
$ \overset{ \eta}{ \Christ}{}^{\mu}{}_{\sigma \mu}( \xi)$, and accordingly the non-metricity tensor $\overset{\textrm{st}}{Q}_{j k \ell}(\xi)$ are given by either $\overset{\theta}{Q}_{j k \ell}(\xi)$ or  $\overset{\eta}{Q}_{j k \ell}(\xi)$.
Note that by changing the parameter $\lambda$ to the affine parameter, say $\tau$, the right hand side of \eqref{arbAPE} vanishes.
Here the affine parameter $\tau$ satisfies $d^2 \tau / d \lambda^2 = f(\lambda) d \tau / d\lambda$.
Then the autoparallel equations become the simple form
\begin{align}
  \frac{d  u^i(\tau)}{d \tau} + \STChrist^i{}_{jk}(\xi) u^j(\tau) u^k(\tau)
  = 0.
\end{align}
For the case of the ST connection $\overset{ \theta}{ \Christ}{}^{i}{}_{j k}( \xi)$, it follows that
\begin{subequations}
\label{thetaAPE}
\begin{empheq}[left=\empheqlbrace]{align}
   &\frac{d^2 \mu}{d \tau} -\frac{4}{\sigma} \frac{d \mu}{d \tau} \frac{d \sigma}{d \tau}= 0, \\
 &\frac{d^2 \sigma}{d \tau} 
 - \frac{3}{\sigma} \left( \frac{d \sigma}{d \tau} \right)^2 =  0.
\end{empheq}
\end{subequations}
The solution curves are $\mu = c_1 \sigma^2 + c_2$ where $c_1$ and $c_2$ are constants. \\
For the case of the ST connection $\overset{ \eta}{ \Christ}{}^{i}{}_{j k}( \xi)$, it follows that
\begin{subequations}
\label{etaAPE}
\begin{empheq}[left=\empheqlbrace]{align}
   &\frac{d^2 \mu}{d \tau} = 0, \\
 &\frac{d^2 \sigma}{d \tau} + \frac{1}{\sigma} \left(\frac{d \mu}{d\tau} \right)^2 
 + \frac{1}{\sigma} \left( \frac{d \sigma}{d \tau} \right)^2 =  0.
\end{empheq}
\end{subequations}
The solution curves are the semi-circles satisfying
\begin{align}
  (\mu - \mu_0)^2 + \sigma^2 = R^2, \quad \sigma > 0,
\end{align}
where $\mu_0$ and $R$ are constants.

Finally we consider the $\theta$-frame, i.e., the coincident gauge of $\overset{\theta}{\Gamma}{}^a{}_{bc}(\theta)=0$.
The components $g_{ij}(\theta)$ of the metric tensor are 
\begin{align}
g_{ij}(\theta) =
\left( \begin{array}{cc}
   -\frac{1}{2 \theta^2} & \frac{\theta^1}{2 (\theta^2)^2} \\[1ex] 
   \frac{\theta^1}{2 (\theta^2)^2}  &  -\frac{(\theta^1)^2}{2 (\theta^2)^3}+\frac{1}{2 (\theta^2)^2}
\end{array}
\right) =
\left( \begin{array}{cc}
   \sigma^2 &  2 \mu \sigma^2  \\[1ex] 
    2 \mu \sigma^2  &  4 \mu^2 \sigma^2 + 2 \sigma^4
\end{array}
\right)
.
\end{align}
Non-zero components of the totally symmetric non-metricity tensor are
\begin{align}
Q_{112}(\theta) &= Q_{121}(\theta) = Q_{211}(\theta) = \frac{\theta^1}{2 (\theta^2)^2} = 2 \sigma^4, \notag \\
Q_{122}(\theta) &= Q_{212}(\theta) = Q_{221}(\theta) = -\frac{\theta^1}{(\theta^2)^3} = 8 \mu \sigma^4,\notag \\
Q_{222}(\theta) &= \frac{3 (\theta^1)^2}{2 (\theta^2)^4} -\frac{1}{(\theta^2)^3} = 24 \mu^2 \sigma^4 + 8 \sigma^6. 
\end{align}

Since $u^2(s)=1$ for the arc-length parameter $s$, which satisfies $ds^2 = g_{ab}(\theta) d\theta^a d \theta^b$, 
the autoparallel equations become
\begin{align}
  \frac{d^2 \theta^a}{d s^2} = -\frac{1}{2} Q_{b c d}(\theta) u^b(s) u^c(s) u^d(s)
   \frac{d \theta^a}{d s}.
\end{align}
Note that the presence  of the term proportional to $d\theta^a / ds$ on the right hand side implies that the arc-length parameter $s$ is not an affine parameter unlike in the Riemannian geometry with metricity.
Naturally, introducing the affine parameter $\tau $ yields $d^2 \theta^a / d \tau^2 = 0$, which leads to the linear solution 
$\theta^a(\tau) = \alpha^a \tau + \beta^a$, where $\alpha^a$ and $\beta^a$ are constants depending on the initial conditions.
However, even in this case, although the components $\theta^a$ change linearly, the basis $\partial_a$ changes point to point due to the non-metricity.

\section{Conclusion}
\label{conclusion}
As stated in Introduction, conventional attempts to synthesize IG with gravitational physics based on GR lead to the conceptual friction, because the dually flat structure in IG do not naturally align with the metric compatible ($Q=0$), torsion-free Riemannian manifold in GR. 
I have reframed IG via the geometric machinery of STG.
Unlike the conventional method in IG, this reframing enables us to distinguish the geometric structure of the $\xi$-coordinates
which is not differential geometrically equivalent to Hessian geometry \cite{Shima}, from that of the $\theta$- or $\eta$-frame. Their relationship is identical to that between the general coordinates and the special coordinates in the coincident gauge in STG.
The $\theta$- or $\eta$-coordinates are the spacial coordinates in the coincident gauge, where all connection coefficients
vanish. On the other hand, in the description using the $\xi $-coordinates, the connection coefficients do not generally vanish, and the Fisher metric is not necessarily Hessian.
In this way, the dually flat connections and the Amari-Centsov tensor emerge as a natural consequence of geometric non-metricity, rather than an arbitrary affine deformation. 
It is revealed that standard information-geometric coordinates emerge naturally within the coincident gauge where the connection vanishes globally.

We hope this work opens up new perspectives in IG.
This research not only provides a new geometric foundation for IG and statistical manifolds but also establishes a robust link between “information dissipation” and “the representation of gravity via non-metricity in STG.”

\section*{Acknowledgments}
The author is deeply grateful to Akio Hosoya for insightful discussions and encouragement.
The author also acknowledges that discussions with Kyosuke Tomonari led to the useful identity
shown in \eqref{formula}.
This work was supported by Japan Society for the Promotion of Science (JSPS) Grants-in-aid for Scientific Research (KAKENHI) grant number 25K0710.

\appendix

\section{The St\"uckelberg  trick}
\label{app1}
The St\"uckelberg  trick \cite{S38} is the procedure to restore diffeomorphism invariance (or general coordinate invariance) to some objects  which appear to violate it, such as in theories of massive gravity \cite{AGS03}.  The St\"uckelberg trick introduces  new scalar fields $ \phi^a (\xi) $ that transform under diffeomorphisms precisely to cancel the symmetry-breaking terms.
Here the scalar fields $\phi^a (\xi)$ are called St\"uckerberg's fields and $a$ is an internal index.

As an example, let us consider the object $\partial_k g_{ij}(\xi)$ on a manifold $\mathcal{M}$
and explain the St\"uckerberg procedure \cite{JK22} restoring the general coordinate invariance to it.
We first remind that under a general coordinate transformation $\xi \to \tilde{ \xi }$, a metric tensor $g_{i j}(\xi)$ transforms as
\begin{align}
g_{ij}(\xi) \to \tilde{g}_{ij}(\tilde{\xi}) = \frac{\partial \xi^{\ell}}{\partial \tilde{\xi}^i} \frac{\partial \xi^m}{\partial \tilde{\xi}^j}  g_{ \ell m}(\xi).
\end{align}
Then the object $\partial_k g_{ij}(\xi)$ transforms as
\begin{align}
 \frac{\partial \tilde{g}_{ij}(\tilde{\xi})}{\partial \tilde{\xi}^k} &= 
\frac{\partial \xi^n}{\partial \tilde{\xi}^k} \frac{\partial}{\partial \xi^n} \left( \frac{\partial \xi^\ell}{\partial \tilde{\xi}^i} \frac{\partial \xi^m}{\partial \tilde{\xi}^j} g_{\ell m}(\xi) \right),
\label{delg}
\end{align}
which does not transforms as a tensor, i.e., it has no general coordinate invariance.

The key idea of the St\"uckelberg procedure to restore the general coordinate invariance is
to perform the corresponding transformation and eventually promote the gauge parameters to the fields.
Now let us apply this to the object $\partial_k g_{ij}(\xi)$ specifically.
First, we introduce $n$ real scalar fields (St\"uckelberg fields) $\{ \phi^a(\xi) \}_{a=1}^n$. 
Note that since they are scalar fields, their values do not change ($\tilde{\phi}^a(\tilde{ \xi }) = \phi^a( \xi )$) under a general coordinate transformation $\xi \to \tilde{\xi}$, but  only their expressions change.
The set of these scalar fields $\phi^a(\xi)$ play a role as a coordinate system in the internal space.

Next we define new object $D_c G_{ab}(\xi)$ by promoting the gauge parameters $\{ \tilde{\xi}^i \}_{i=1}^n$ in \eqref{delg} to the St\"uckelberg fields $ \{ \phi^a(\xi) \}_{a=1}^n$, i.e.,
\begin{align}
  D_c G_{ab}(\xi) :=  \frac{\partial \xi^k}{\partial \phi^c}\frac{\partial}{\partial \xi^k} \left( \frac{\partial \xi^i}{\partial \phi^a}    \frac{\partial \xi^j}{\partial \phi^b} \; g_{ij}(\xi) \right),
 \end{align}
 where
 \begin{align}
 G_{ab}(\xi) :=   \frac{\partial \xi^i}{\partial \phi^a}    \frac{\partial \xi^j}{\partial \phi^b} \; g_{ij}(\xi), 
 \label{Gab}
 \end{align}
and the new derivative $D_c := \frac{\partial \xi^k}{\partial \phi^c}\frac{\partial}{\partial \xi^k}$.
Note that $G_{ab}(\xi)$ transforms as a scalar as follows.
\begin{align*}
      \tilde{G}_{ab}( \tilde{\xi}) &=
      \left( \frac{\partial \tilde{\xi}^k}{\partial \xi^m} \frac{\partial \xi^m}{\partial \phi^a}\right)    \left( \frac{\partial \tilde{\xi}^{\ell}}{\partial \xi^n} \frac{\partial \xi^n}{\partial \phi^b}\right)  \left( \frac{\partial \xi^i}{\partial \tilde{\xi}^k}  \frac{\partial \xi^j}{\partial \tilde{\xi}^{\ell}} g_{ij}(\xi)\right) 
      = \underbrace{  \frac{\partial \xi^i}{\partial \tilde{\xi}^k}  \frac{\partial \tilde{\xi}^k}{\partial \xi^m} }_{\delta^i_m}
      \underbrace{ \frac{\partial \xi^j}{\partial \tilde{\xi}^{\ell}}  \frac{\partial \tilde{\xi}^{\ell}}{\partial \xi^n}}_{\delta^j_n} 
      \frac{\partial \xi^m}{\partial \phi^a}    \frac{\partial \xi^n}{\partial \phi^b} g_{ij}(\xi) \notag \\
      &= \frac{\partial \xi^i}{\partial \phi^a}    \frac{\partial \xi^j}{\partial \phi^b} g_{ij}(\xi) = G_{ab}(\xi). 
\end{align*}
Then we readily  see that the new object $D_c G_{ab}(\xi)$ also transforms as a scalar. Consequently $D_c G_{ab}(\xi)$ has the general coordinate transformation invariance.

The co-basis $\{ d\xi^i \}$ on $\mathcal{M}$ and the co-basis $\{ d\phi^a \}$ in the internal space (or moving frame) are related by
\begin{align}
   d\xi^i = \frac{\partial \xi^i}{\partial \phi^a} d \phi^a.
\end{align}
It is worthwhile to note that an internal index $a$ is not a tensor index on $\mathcal{M}$.
Note that the quantity $\partial \xi^i / \partial \phi^a$ transforms as a contravariant vector under the general coordinate transformation $\xi \to \tilde{\xi}$ on $\mathcal{M}$.
Since the components $(\partial \xi^i / \partial \phi^a) (\partial \xi^j / \partial \phi^b) $ in \eqref{Gab} transform as a second-order contravariant tensor and the components $g_{ij}(\xi)$ of a metric tensor $g$ transform as a second-order covariant tensor, they completely cancel each other out under the general coordinate transformation. This is a key trick in St\"uckelberg's procedure.

\section{Useful identity}
\label{app2}
Here we provide a simple proof of the useful identity 
\begin{align}
Q_{ijk} &- Q_{kji} = T_{jik}+ T^{\star}_{jik},
\label{formula}
\end{align}
which relates the non-metricity $Q_{ijk}$, the torsion $T^k{}_{ij} := \Gamma^k{}_{ij} - \Gamma^k{}_{ji}$ and the dual torsion $T^{\star}{}^k{}_{ij} := \Gamma^{\star}{}^k{}_{ij} - \Gamma^{\star}{}^k{}_{ji}$.

By writing down the definition of non-metricity \eqref{Qcf}, we have
\begin{align*}
Q_{ijk} &- Q_{kji} = \nabla_i g_{jk} - \nabla_k g_{ji} 
= \underbrace{\partial_i g_{jk} - \Gamma^{\ell}{}_{ij} g_{\ell k}}_{\Gamma^{\star}_{jik}} 
- \underbrace{ \Gamma^{\ell}{}_{ik}  g_{j \ell}}_{\Gamma_{jik}}
\underbrace{-\partial_k g_{ji} + \Gamma^{\ell}{}_{kj} g_{\ell i}}_{-\Gamma^{\star}_{jki}} +\underbrace{\Gamma^{\ell}{}_{ki}  g_{j \ell}}_{\Gamma_{jki}} \notag \\
&= \Gamma_{jik}-\Gamma_{jki} + \Gamma^{\star}_{jik} - \Gamma^{\star}_{jki} 
= T_{jik}+ T^{\star}_{jik},
\end{align*}
where we used \eqref{dualC}.

\end{document}